\documentclass[12pt]{article}
\usepackage {amsmath}
\usepackage[dvips]{graphicx}
\usepackage{amssymb}
\usepackage{cite}
\newcommand{\be}{\begin{equation}}
\newcommand{\ee}{\end{equation}}
\newcommand{\bear}{\begin{eqnarray}}
\newcommand{\eear}{\end{eqnarray}}

\newcommand{\lapproxeq}{\lower .7ex\hbox{$\;\stackrel{\textstyle  
<}{\sim}\;$}} 
\newcommand{\gapproxeq}{\lower .7ex\hbox{$\;\stackrel{\textstyle  
>}{\sim}\;$}} 
\newcommand{\stackdown}[2]{\lower 1.4ex\hbox{$\;\stackrel{\textstyle{#1}}  
{\scriptstyle{#2}}\;$}}
\newcommand{\beq}{\begin{equation}} 
\newcommand{\eeq}{\end{equation}} 

\newcommand{\ba}{\begin{eqnarray}}
\newcommand{\ea}{\end{eqnarray}}

\newcommand{\bea}{\begin{eqnarray}}
\newcommand{\eea}{\end{eqnarray}}

\makeatletter 
\def\slash{\@ifnextchar[{\fmsl@sh}{\fmsl@sh[0mu]}} 
\def\fmsl@sh[#1]#2{%
  \mathchoice 
    {\@fmsl@sh\displaystyle{#1}{#2}}%
    {\@fmsl@sh\textstyle{#1}{#2}}%
    {\@fmsl@sh\scriptstyle{#1}{#2}}%
    {\@fmsl@sh\scriptscriptstyle{#1}{#2}}} 
\def\@fmsl@sh#1#2#3{\m@th\ooalign{$\hfil#1\mkern#2/\hfil$\crcr$#1#3$}} 
\makeatother 
\begin{document} 
\begin{titlepage}  
\begin{flushright} 
\parbox{4.6cm}{hep-th/0402228\\
               UA-NPPS/BSM-04/01 }
\end{flushright} 
\vspace*{5mm} 
\begin{center} 
{\large{\textbf {On the brane coupling of
Unified orbifolds with gauge interactions in the bulk
}}}\\
\vspace{14mm} 
{\bf G. A.~\ Diamandis}, \, {\bf B. C.~\ Georgalas}, \, 
{\bf P.~\ Kouroumalou } and \\ {\bf A. B.~\ Lahanas} 

\vspace*{6mm} 
  {\it University of Athens, Physics Department,  
Nuclear and Particle Physics Section,\\  
GR--15771  Athens, Greece}

\end{center} 
\vspace*{25mm} 
\begin{abstract}
In the on-shell formulation of $D=5\,$, $N=2$ supergravity, compactified
on $S^1/Z_2$, we extend the results of Mirabelli and Peskin describing the 
interaction of the bulk fields with matter which is assumed to be confined on the 
brane. The novel characteristics of this approach are : 
Propagation of both gravity and gauge fields in the bulk, 
which offers an alternative for a unified description of
models in extra dimensions and 
use of the on-shell formulation avoiding the complexity
of off-shell schemes which involve numerous auxiliary fields. 
We also allow for nontrivial superpotential interactions of the chiral matter fields. 

The method we employ uses the N{\" o}ther procedure and our findings are
useful for building models advocating propagation of the
gauge degrees of freedom in the bulk, in addition to gravity.
\end{abstract} 
\end{titlepage} 
\newpage 
\baselineskip=18pt 
\section{Introduction}

It is well established that the Standard Model (SM)
 describes succesfully all particle interactions at low energies.
 On the other hand it is understood that SM is an
 effective theory. At  high energies, 
 description of the elementary particle interactions demands
 a generalization of the SM. Assuming a unified
 description
 in terms
 of a renormalizable field theory, up to very high
 energies lead to favorable generalization namely GUT theories
  \cite{gut1}, among which supersymmetric GUTs \cite{gut2}
  play a central r\^ole.
 Consistent inclusion of gravity dictates that these
 generalizations should be effective descriptions
 of a more fundamental underlying theory.
 String Theory \cite{string}
 is the most prominent candidate
 for this aim. Indeed from the 10 dimensional
 field theory, which is the effective point limit
 of the String Theory  we can get, by suitable
 compactifications of the extra dimensions, consistent
 four-dimensional models compatible with the SM
 \cite{dienes1}.
 Along these lines it has been conjectured that
 one or two dimensions  may be compactified
 at different scales, lower from the remaining ones
  \cite{large1}. Also after the developments
 concerning the
 duality symmetries of String Theory and in the
 framework of M-Theory \cite{wittenm, bachasm}, the
 idea that our world may be a
 brane embedded in a higher dimensional space has
 recently
 attracted much interest and has been studied intensively
 \cite{horava, lykken, large2,  ovrut, antoniadis1}.
 Besides the original compactifications new
 possibilities have been proposed \cite{randall1}. It has
 been also
 recognized
 that String/M theory may lead to brane-world models in
 which one of the extra dimensions can be even non-compact
 \cite{randall2, rubakov}.
 In all these models the four-dimensional world is
 a brane, on
 which the matter fields live, while gravity, and in some interesting
 cases
 the gauge and the Higgs fields, 
 propagate also in the transverse extra dimensions of the bulk space.

 In the majority of the cases studied, in an attempt
 to build realistic models,
 the bulk is a
 five-dimensional space \cite{peskin}.
 In these models the corresponding
 backgrounds may be of Minkowski or Anti-de-Sitter type.
 Effects of the above consideration in
 specific 
 GUT models have been
 also considered. The assumed background for these models is of
 Minkowski type and questions regarding the unification 
 and supersymmetry breaking scales
 have been addressed to. In this direction
 assuming the fifth dimension very large, of the
 TeV scale, non supersymmetric extensions of the SM even
 without the need of unification have been considered
 \cite{dudas, dudas2, kiritsis, kawamura, altarelli, russell}.
 On the other
 hand models embedding the SM in an Anti-de-Sitter five
 dimensional space have been also discussed
 \cite{pomarol1, pomarol}.

 In view of the aforementioned developments the study of the
 five dimensional supergravities has been
 revived \cite{fivesugra, gunaidin, recent, fre}. This is quite
 natural since after all gravity is in the center
 of all these attempts and it is legitimate to
 assume that we have to treat the fifth
 dimension before going to a "flat limit". In these recent
 considerations of the five-dimensional supergravity
 no specific model based on a particular gauge group has been introduced
 so far. Also the interaction of the brane multiplets with the bulk gauge fields, 
essential for Supersymmetry and the transmition of its breaking, \cite{peskin}, 
has not been studied in the context of D=5, N=2 supergravity models 
in which gauge fields are allowed to propagate in the bulk in addition to gravity.  
In this note we undertake this in the on shell formulation of five - dimensional supergravity.

\section{Setting the model}

We consider a five-dimensional Yang-Mills supergravity model.
  The field content of the model
is \cite{fivesugra}
 \be
 \{e_{\tilde{\mu}}^{\tilde{m}},\Psi_{\tilde{\mu}}^{i}, A_{\tilde{\mu}}^{I}, 
 \lambda^{i a}, \phi^{x} \}
 \label{multiplets}
 \ee
where $\tilde{\mu}=(\mu,\,5)$ are curved and $\tilde{m}= (m,\dot{5})$ are flat
five-dimensional indices, with $\mu$, $m$  their corresponding four 
dimensional indices. The remaining indices are 
$\;I = 0,1,\ldots, n \;$, $\;a = 1,\ldots, n \;$ and $\;x = 1,\ldots, n \;$. 
  The supergravity multiplet consists
 of the f\"{u}nfbein $e_{\tilde{\mu}}^{\tilde{m}}$, two gravitini
 $\Psi_{\tilde{\mu}}^{i}$ and the graviphoton $A_{\tilde{\mu}}^{0}$, where
 $i=1,2$ is the symplectic
 $SU(2)_{R}$ index. Moreover, there exist n
 vector multiplets,  counting the
 Yang-Mills  fields ($A_{\tilde{\mu}}^a$).  
 The spinor and the scalar fields included in the vector
  multiplets are collectively denoted by
 $\lambda^{i a}$, $\phi^{x}$ respectively.
 The indices $a$, $x$ are flat and curved
 indices respectively of the $n$-dimensional manifold
 $\mathcal{M}$ parametrized by the scalar fields.
 This manifold is embedded
 in an $(n + 1)$-dimensional space and it is determined
 by the cubic constraint
 \be
 C_{I J K}
 h^{I} h^{J} h^{K}=1.
 \label{ccon}
 \ee
 $h^{I}$ are functions of the scalar fields
 defining the embedding of the manifold $\mathcal{M}$.
 $C_{I J K}$ are
 constants symmetric in the three indices.

 The assumption that the gauge interactions propagate in the five-dimensional 
bulk while only the matter fields are localized on the branes is implemented as follows.
We consider five-dimensional vector fields $A^{a}_{\tilde{\mu}}$, 
and we perform the $S_1/Z_2$ orbifold by assigning $Z_2$-even parity to the four dimensional
part of the vectors $A^a_{\mu}$ and $Z_2$-odd parity to the fifth component $A^a_{5}$.
 The five-dimensional Yang-Mills Einstein 
Lagrangian has to be even for the parity assignments to be consistent. The only
term that may cause problem is the Chern-Simons term 
\ba
{\frac
{\epsilon ^{\tilde{\mu} \tilde{\nu} \tilde{\rho} \tilde{\sigma} \tilde{\lambda}}
}{6 \sqrt{6}}}
C_{IJK}
  \Big{\{} F^{I}_{\tilde{\mu} \tilde{\nu}} F^{J}_{\tilde{\rho} \tilde{\sigma}} 
A_{\tilde{\lambda}}^{K} +
 \frac{3}{2} g F^{I}_{\tilde{\mu} \tilde{\nu}} A_{\tilde{\rho}}^{J}
 (f_{LF}^{K} A_{\tilde{\sigma}}^{L} A_{\tilde{\lambda}}^{F}) 
 + \frac{3}{5} g^2 (f_{GH}^{J} A_{\tilde{\nu}}^{G} A_{\tilde{\rho}}^{H})
 (f_{LF}^{K} A_{\tilde{\sigma}}^{L} A_{\tilde{\lambda}}^{F}) A_{\tilde{\mu}}^{I} \Big{\}} 
\label{chern}
\ea
where $f_{IJ}^{K}$ above are the structure constants. 
In particular $f_{ab}^{c}$ are the
structure constants of the non-abelian gauge group and $f_{IJ}^{K}=0$
if any one of the indices is $0$. $g$ is the gauge coupling constant.
This term is in general odd for the assignment given above. Nevertheless if we chose
the coefficients of the cubic constraint to be in the canonical basis
\be
C_{000}=1, \quad C_{0ab}=-\frac{1}{2} \delta_{ab}, \quad C_{00a}=0, 
\quad C_{abc}={\mathrm{arbitrary}}
\label{solcon}
\ee
and give $Z_2$-odd parity to the four dimensional part of the graviphoton, $A^0_{\mu}$,
and $Z_2$-even parity to its fifth component $A^0_{5}$ then the Chern-Simons term becomes
even and the parity assignment is consistent with the choice $\; C_{abc}=0\;$ which we assume in the following.
For example one of the terms in (\ref{chern}) is 
$ \sim \epsilon ^{\mu \nu \rho \sigma 5} 
C_{IJK} F^{I}_{\mu \nu} F^{J}_{\rho \sigma} A_{5}^{K} \;$ and 
if we take $A_{\mu}^{a}$ to be $Z_2$-even and $A_{5}^{a}$ 
to be $Z_2$-odd, obviously $C_{abc}$ have to vanish. Also since in the
canonical basis we are enforced to take $C_{0ab}=-\frac{1}{2} \delta_{ab}$ 
the corresponding Chern-Simons term is even with 
$A^0_{\mu}$   $Z_2$-odd and $A^0_{5}$  $Z_2$-even. 
The remaining terms in the expansion of (\ref{chern}) may be treated accordingly.

Considering now the full spectrum, the $Z_2$-even fields are
$$
e_{\mu}^m, \quad e_5 ^{\dot{5}}, \quad \Psi_{\mu}^1, \quad \Psi_{5}^2, 
\quad A_{5}^0, \quad A_{\mu}^a, \quad \lambda^{1 a},
$$
while the $Z_2$-odd fields are
$$
e_{\mu}^{\dot{5}}, \quad e_5^m, \quad \Psi_{\mu}^2, \quad \Psi_{5}^1, 
\quad A_{\mu}^0, \quad A_{5}^a, \quad \lambda^{2 a}, \quad \phi ^{x}.
$$ 

These parity assignments complete the $S_1/Z_2$ orbifold. All the fields of the
abovementioned spectrum propagate in the bulk. Only the even fields propagate on
the two branes located at $x^{5}=0$ and $x^{5}=\pi R$, that is the fixed points of the
$Z_2$ transformation. 

The spectrum of the even fields respects four-dimensional N=1 supersymmetry. The model is supplied
by chiral multiplets localized on the branes. The coupling of these multiplets to the bulk fields
is determined by the N=1 Supersymmetry invariance of the total action. 

 The bulk Lagrangian is \cite{gunaidin}
 \ba
 {\mathcal{L}}_{0}/e^{(5)}  &=& -\frac{1}{2} R^{(5)} + \frac{i}{2} \bar{\Psi}_{i 
\tilde{\mu}}
\gamma ^{\tilde{\mu} \tilde{\nu} \tilde{\rho}} \nabla _{\tilde{\nu}} \Psi_{\tilde{\rho}} ^{i}
- \frac{1}{4}
 \mathring{a}_{I J}
 F_{\tilde{\mu} \tilde{\nu}}^{I}
 F^{I \, \tilde{\mu} \tilde{\nu}} 
-  \frac{1}{2} g_{xy}  
 ({\mathcal{D}}_{\tilde{\mu}} \phi ^ {x})
 ({\mathcal{D}}^{\tilde{\mu}} \phi ^ {y}) 
\nonumber \\ 
&+& \mathrm{Fermion } \;+\;  \mathrm{\;Chern - Simons \; \; terms } 
\label{action}
\ea

The tensor $\mathring{a}_{I J}$, appearing in the kinetic terms 
of the gauge fields, is the
restriction of the metric of the $(n +1)$-
dimensional space on the $n$-dimensional manifold
of the scalar fields given by 
\be
\mathring{a}_{IJ} =
 -2 C_{I J K}h^{K} +
 3 h_{I} h_{J}
\label{alpha} \; \;,
\ee
where 
$\; 
h_{I} = C_{IJK}
h^{J}h^{K}=
\mathring{a}_{IJ} h^{J}
\;$
and $\;g_{xy}= h_{x}^{I}
h_{y}^{J} \mathring{a}_{IJ} \;$ 
is the metric of the
$n$-dimensional manifold ${\mathcal{M}}$. 
In these equations  $h_{x}^{I}=-\sqrt{\frac{3}{2}}
h^{I},_{x} $ and   $h_{Ix} = \sqrt{\frac{3}{2}}
 h_{I},_{x}$. Note also that the following
relations hold
\be
h^{I} h_{I}=1, \quad
 h_{x}^{I} h_{I}=h^{I}
 h_{Ix}=0 \; \; .
\label{relations}
 \ee
 
With the parity assignments we have adopted, $h^0$ is even, while
$h^x=\phi ^x$ are odd. Furthermore on the fixed points where the odd quantities vanish,
$h^0=1$. Analogous relations hold for the $h_I$'s.

\section{The Supersymmetry Transformations}
Recalling the linearized supersymmetry transformations
 of the bulk fields
\ba
\delta e^{\tilde{m}}_{\tilde{\mu}} &=& 
i \bar{\epsilon}_{i}
\gamma^{\tilde{m}} \Psi_{\tilde{\mu}}^{i} \nonumber \\
\delta \Psi_{\tilde{\mu}}^{i} &=& 2 \nabla _{\tilde{\mu}}
(\omega) \epsilon^{i}
- \frac{h_I}{2\sqrt{6}}\gamma_{\tilde{\mu}}^ {\tilde{\nu}
 \tilde{\rho}}
F_{\tilde{\nu} \tilde{\rho}}^I \epsilon^{i} -
\frac{2 h_I}{\sqrt{6}}\gamma^{\tilde{\rho}} F_{\tilde{\mu} 
\tilde{\rho}}^I \epsilon^{i} \nonumber \\
\delta A_{\tilde{\mu}}^I &=& -i h_a^I \bar{\epsilon}_{i}
\gamma _{\tilde{\mu}} \lambda ^{a i}- \frac{i \sqrt{6}}{2}
h^I \bar{\Psi}_{\tilde{\mu} i} \epsilon^{i} \nonumber \\
\delta \lambda ^{a i} &=& - f^a _x \gamma ^{\tilde{\mu}}
D_{\tilde{\mu}} \phi ^x \epsilon^{i} - \frac{1}{2}
h ^a _I \gamma ^{\tilde{\mu} \tilde{\nu}} \epsilon^{i} 
F_ {\tilde{\mu} \tilde{\nu}}^I \nonumber \\
\delta \phi ^x &=& -i f_a ^x \bar{\epsilon}_{i}
\lambda ^{a i}
\label{fivedimtr} 
\ea
we see that the parity assignments are consistent if  
the supersymmetry parameters 
\be
 \epsilon ^1 = \left( 
 \begin{array}{c}
 \varepsilon \\
 \bar{\zeta}
 \end{array} \right ), \,
\epsilon ^2 = \left( 
 \begin{array}{c}
 \zeta\\
 -\bar{\varepsilon} 
 \end{array} \right )
\ee
is taken to consist of even $\varepsilon  $ and odd $\zeta$. Recall that 
$ \epsilon ^1, \epsilon ^2$ are symplectic Majorana.  
It is easy to see that under $\varepsilon$ transformations the
fields of the Radion supermultiplet, 
$ \left \{ \frac{1}{\sqrt{2}} h^0 e_5^{\dot{5}}  +i \sqrt{\frac{1}{3}} A_{5}^0
, \quad h^0 \Psi_{5}^2 \right \}$ 
transform like a chiral multiplet,
while the transformation of the even fields under $\varepsilon$-supersymmetry reads
\ba
\delta e^m_{\mu} &=& i \left( \varepsilon  \sigma ^m \bar{\psi}_{\mu} +
\bar{\varepsilon } \bar{\sigma} ^m \psi_{\mu} \right) \nonumber \\
\delta \psi_{\mu} &=& 2 \nabla _{\mu} (\omega) \varepsilon  + 
\frac{i \; h_0}{2 \sqrt{6}} \left[(\sigma_{\mu} \bar{\sigma}^{\nu}
 - \sigma^{\nu} \bar{\sigma} _{\mu}) + 4\delta _\mu^\nu  \right]
\varepsilon F^0_{\nu {5}}\;e_{\dot{5}}^5 + ...
 \nonumber \\
\delta A_{\mu}^I &=& -i h_{a}^I  
\left( \varepsilon  \sigma_{\mu} \bar{\lambda}^a +
\bar{\varepsilon } \bar{\sigma}_{\mu} \lambda^a \right) + ...  \nonumber \\
\delta \lambda^a &=&  -h^{a}_I \sigma ^{\mu \nu} \varepsilon  F_{\mu \nu}^I
+i f_x ^a \partial _{5} \phi ^x e_{\dot{5}}^5 \varepsilon + ... 
\label{eventrans}
\ea
where $\psi_\mu \equiv \Psi_{\mu \, L}^1$ and $ \lambda^a \equiv \lambda^{1a}_{L}$. 
The ellipsis in eq. (\ref{eventrans}) stand for even products of odd fields, and hence vanishing  on the brane.   
Thus we see that on the branes determined by the orbifold construction, we get the 
four-dimensional $N=1$ on-shell transformation of the supergravity multiplet, the Yang-Mills
vector multiplets and one chiral multiplet, the radion multiplet, surviving from 
the five-dimensional supergravity
multiplet. Notice however the  appearance of an extra $\partial _{5} \phi ^x $ term, and a 
$F^0_{\nu 5}$ dependent term  in the gaugino and gravitino transformation laws respectively.

\section{Bulk Gravity and Gauge Couplings of the Brane Multiplets}
The matter fields are considered to be localized on the branes at the fixed points  $x^{5}=0$ and
$x^{5}=\pi R$. 
For the purposes of this work it suffices to consider only the 
brane at $x^{5}=0$. The treatment of fields living on the brane at $x^{5}=\pi R$ is  
done similarly.

The requirement of $N=1$ local supersymmetry invariance on the branes
determines the on-shell 
couplings of these fields to the gravity and gauge multiplets. 
These can be found following N\"{o}ther's procedure.
This procedure is used in the on-shell formulation of local supersymmetry 
\cite{newvan}, where the r{\^o}le of the gauge field is played by the
 gravitino, while the gauge current is the supercurrent. However in the case 
of supersymmetry besides the modification of the Lagrangian the transformation
 laws should be also modified accordingly. This is well understood since the 
on-shell formulation follows from the off-shell after eliminating the 
auxiliary fields by solving the equations of motion which are modified upon 
changing the Lagrangian at each step. 

The original Lagrangian is
\be
\mathcal{L}_{\mathrm{orig}} = \mathcal{L}_0 + \mathcal{L}_b
\label{orig}
\ee
with $\mathcal{L}_b$ the "brane" part including the interactions of the matter
 fields, localized on the brane, with the "projections" of the bulk fields, 
gravity and gauge fields, on the brane. The original SUSY transformations will 
be denoted by $\delta _0$. $\mathcal{L}_0$ is invariant under $ \delta _0 $, i.e. 
$\delta _0  \mathcal{L}_0 = 0$, but not  $\mathcal{L}_b $ that is 
 $\delta _0  \mathcal{L}_b \neq 0$. 
As already stated we must modify the original theory by adding new terms, 
$\Delta \mathcal{L} $, so that the total Lagrangian  
\[
\mathcal{L}_s = \mathcal{L}_0 + \mathcal{L}_b + \sum \limits_k \Delta_k \mathcal{L} 
\]
is invariant under the modified SUSY transformations denoted by $\delta _s$,
\[
\delta_s  = \delta_0  + \sum \limits_k \delta_k 
\]
that is $\delta_s \mathcal{L}_s = 0$. We will proceed iteratively and 
in the above sums  $k$ denotes the iteration step. 
 
In order to derive the gravitational couplings we ignore for the 
moment the gauge interactions and consider for 
simplicity just one chiral multiplet on the brane at $x^5=0$.
Thus we start from $\mathcal{L}_b$ which for one chiral multiplet,  
$(\varphi , \chi )$, has the form
 \footnote{It is usefull to define 
$\; \Delta_{(5)}(x^5) \; \equiv\;  \delta (x^5)/ e_5^{\dot{5}} = e_{\dot{5}}^5 \delta (x^5) \;$. 
Note that with the parity assignments we have adopted
$\; e^{(5)}\;\Delta_{(5)}\;=\;e^{(4)}\; \delta (x^5) \;$  on the brane.}
\be
\mathcal{L}_b = - {e^{(5)}} \Delta_{(5)}   \left(
\partial _{\mu} \varphi \partial ^{\mu} \varphi ^{*} 
+ i \bar{\chi } \bar{\sigma}^{\mu}D _{\mu}\chi 
\right).
\label{brane1}
\ee
In order to facilitate the discussion we have ignored at this stage 
superpotential and gauge interactions.
Since $ \mathcal{L}_b $ includes dependencies on the vierbein it facilitates
to write 
\be
\delta _0  = \delta _0 ^{(e)} + \delta _0 ^{(\mathrm{rest})} 
\ee
with $\delta_0^{(e)} $, $\delta_0^{(\mathrm{rest})} $  denoting variations acting 
on the vierbein and the remaining fields respectively. 
With this we get from (\ref{brane1})
\be
\delta_0 \mathcal{L} _b  = \delta_0^{(e)} 
\left[-
 {e^{(5)}} \Delta_{(5)}   \left(
\partial _{\mu} \varphi \partial ^{\mu} \varphi ^{*} 
+i \bar{\chi } \bar{\sigma}^{\mu}\partial_{\mu}\chi 
\right)
\right] + {e^{(5)}} \Delta_{(5)}   
 \left( J_{\mu} \partial ^{\mu}\varepsilon + h.c. \right) 
\label{delta1}
\ee
where $J^{\mu} $ in (\ref{delta1}) is the (N{\" o}ther) supercurrent
given by
$\;J^{\mu} = \sqrt{2}
\chi \sigma^{\mu} \bar{\sigma}^{\nu} \partial_{\nu} \varphi ^{*} \;$. 
According to N{\" o}ther's procedure in order to eliminate the last term in 
(\ref{delta1}) we must add a term $ \Delta _1 \mathcal{L} $ 
while no change of SUSY transformations is required at
this stage. Thus
\bea
\delta _1 = 0  \; \; ,\; \;
\Delta _1 \mathcal{L} = -\frac{1}{2 } {e^{(5)}} \Delta_{(5)}  
\left(J_{\mu} \psi ^{\mu}+ h.c. \right)
\label{first}
\eea
since in our conventions $\delta _0 \psi_{\mu} \sim  2D_{\mu}\varepsilon + ...$.
Next we have to check the invariance of the so constructed Lagrangian
$\mathcal{L}_0 + \mathcal{L}_b + \Delta _1 \mathcal{L} $ and modify it 
accordingly if it happens to be non-invariant under the new SUSY transformation 
law $\delta_s  = \delta_0  +\delta_1$, which however at this stage, due to the vanishing of 
$\delta _1 $ coincides with the original transformation  $\delta _0 $.
Using the gravitino transformation law given by the second equation in 
eq. (\ref{eventrans}) one gets 
\bea
& &\delta_s \left
( \mathcal{L} _0 + \mathcal{L} _b + \Delta _1\mathcal{L} \right) = \nonumber \\
& & \delta_0^{(e)} 
\left[-
{e^{(5)}} \Delta_{(5)}   \left(
\partial _{\mu} \varphi \partial ^{\mu} \varphi ^{*} 
+i \bar{\chi } \bar{\sigma}^{\mu}\partial _{\mu}\chi 
\right)
\right]
- \frac{1}{2} {e^{(5)}} \Delta_{(5)}   
\left[\left( 
\delta _0J_{\mu} \right) \psi ^{\mu} + h.c \right]
\nonumber \\
& & + \frac{1}{\sqrt{6}} {e^{(5)}} \Delta_{(5)}   \;F_{\mu \dot{5}}^0
\;
\delta _0 \left(  J^{(\varphi)  \, \mu} - \frac{1}{2}
J^{(\chi) \, \mu}  \right) 
+ \delta _0 ^{(e)} \left( -\frac{1}{2} {e^{(5)}} \Delta_{(5)}   
  J_{\mu} \psi ^{\mu} +  h.c. \right) 
\label{dtotal11}
\eea
The third term in (\ref{dtotal11}) follows from the 
second term of the
gravitino transformation in (\ref{eventrans}), as can be verified by  
a straightforward algebra,  
and $J^{(\varphi) }$ , $J^{(\chi)} $ denoting the
$U_R(1)$ currents of $\varphi $ and $\chi $ fields, given
 by
\be
J_{\mu}^{(\varphi) }= -i \varphi ^{*} 
\stackrel{\leftrightarrow }{\partial }_{\mu} \varphi \,\,, \quad
J_{\mu}^{(\chi) }=  \chi \sigma_{\mu} {\bar \chi }
\ee
Note
also that 
we have not included the spin connection $\omega _{\mu \, mn}$ for 
lack of space. Its contribution at each stage
is determined by fully covariantizing the results.

The first three terms in (\ref{dtotal11}) are cancelled if we 
modify the SUSY transformations as it appears below
\bea
\delta _2 \; \varphi  &=& 0 , \quad
\delta _2 \chi \;  = -i \sigma^{\mu} \bar{\varepsilon} \left(\psi_{\mu} \chi 
\right) \nonumber \\
\delta _2 e_{\mu}^m   &=& 0 , \quad
\delta _2  \psi_{\mu} =  \frac{i}{2} \Delta_{(5)}   
 \left(  J_{\mu}^{(\varphi )}\varepsilon  
- \sigma_{\mu \nu} \varepsilon J^{(\chi )\, \nu} \right) 
\label{coco}     
\eea
and add a term to the Lagrangian given by
\bea
\Delta _2 \mathcal{L} &=&  {e^{(5)}} \Delta_{(5)}  \Big \{   \;
\frac{1}{4} 
J_{\sigma}^{(\chi )}
\left[ \; i E^{\mu \nu \rho \sigma}
\left(\psi_{\mu} \sigma _{\nu} \bar{\psi}_{\rho} \right) + 
\left( \psi_{\mu} \sigma^{\sigma} \bar{\psi}^{\mu} \right)  \right]  \nonumber \\
&-& \frac{i}{4} \;
E^{\mu \nu \rho \sigma}J_{\sigma}^{(\varphi)}
\psi_{\mu} \sigma_{\nu} \bar{\psi}_{\rho} 
 - \frac{1}{\sqrt{6}}
\Big(  J^{(\varphi ) \, \mu} -
 \frac{1}{2} J^{(\chi  ) \, \mu}   \Big) F_{\mu \dot{5}}^0 \; \Big \}  
 \; ,
 \label{second}
\eea
where $E^{\mu \nu \rho \sigma}$ is the four dimensional antisymmetric tensor.
The 
last term in (\ref{second}) is needed for the cancellation of the 
$F_{\mu \dot{5}}^0 \left( - J^{(\varphi) \, \mu} + \cdots \right)$ 
term in (\ref{dtotal11}). The need of introducing the remaining terms 
will be clarified in the following.

We next have to check the invariance of $ \mathcal{L}_0 + \mathcal{L}_b 
+ \Delta _1 \mathcal{L} + \Delta _2 \mathcal{L} $ under 
$\delta _s$ transformations.
Since $\delta _1 = 0$ we have
\be
\delta _s( \mathcal{L}_0 + \mathcal{L}_b 
+ \Delta _1 \mathcal{L} + \Delta _2 \mathcal{L} ) = 
\delta _0( \mathcal{L}_0 + \mathcal{L}_b 
+ \Delta _1 \mathcal{L}  ) +
\delta _0( \Delta _2 \mathcal{L}) + 
\delta _2( \mathcal{L}_0 + \mathcal{L}_b 
+ \Delta _1 \mathcal{L} + \Delta _2 \mathcal{L} )
\label{dtotal3}
\ee
As we have already discussed, from the transformation 
$\delta _0( \mathcal{L}_0 + \mathcal{L}_b 
+ \Delta _1 \mathcal{L}  ) +
\delta _0( \Delta _2 \mathcal{L})$ only the term
$\delta _0^{(e)} \left( \frac{e^{(5)}}{e_5^{\dot{5}}}  J_{\mu} \psi ^{\mu} + h.c.\right)$ 
survives. 
In fact the variation $\delta _0( \Delta _2 \mathcal{L}) $ is given by
\begin{align}
\delta _0( \Delta _2 \mathcal{L}) =&
  \frac{1}{\sqrt{6}}
\left [ - \delta _0 \left( {e^{(5)}} \Delta_{(5)}   F_{\mu \dot{5}}^0 \right)
 \Big(  J^{(\varphi ) \, \mu}  -
 \frac{1}{2} J^{(\chi  ) \, \mu}   \Big) - 
{e^{(5)}} \Delta_{(5)}  F_{\mu \dot{5}}^0 \delta_ 0  
 \Big(  J^{(\varphi ) \, \mu} -
\frac{1}{2}J^{(\chi  ) \, \mu}   \Big) \right] 
         \nonumber \\
&+  \delta _0^{(e)} \left \{ \frac{1}{4} {e^{(5)}} \Delta_{(5)}   
 \left[i E^{\mu \nu \rho \sigma}
\left(\psi_{\mu} \sigma _{\nu} \bar{\psi}_{\rho} \right)
 \left( J_{\sigma}^{(\chi )} - J_{\sigma}^{(\varphi )} \right)
+
\left( \psi_{\mu} \sigma^{\sigma} \bar{\psi}^{\mu} \right)
J_{\sigma}^{(\chi )}   \right]  \right \}
       \nonumber \\
&+    \frac{1}{4}  {e^{(5)}} \Delta_{(5)}   
\left[i E^{\mu \nu \rho \sigma}
\left(\psi_{\mu} \sigma _{\nu} \bar{\psi}_{\rho} \right)
 \delta _0 
\left( J_{\sigma}^{(\chi )} - J_{\sigma}^{(\varphi )} \right)
 + 
\left( \psi_{\mu} \sigma^{\sigma} \bar{\psi}^{\mu} \right)
\delta _0 J_{\sigma}^{(\chi )}   \right]         \nonumber \\
&+ \frac{1}{4}  {e^{(5)}} \Delta_{(5)}   
\left[i  E^{\mu \nu \rho \sigma}
\left( J_{\sigma}^{(\chi )} - J_{\sigma}^{(\varphi )}\right) 
\delta _0\left(\psi_{\mu} \sigma _{\nu} \bar{\psi}_{\rho} \right)
 + 
 J_{\sigma}^{(\chi )}   
\delta _0 \left( \psi_{\mu} \sigma^{\sigma} \bar{\psi}^{\mu} \right)
\right]  
\label{dtotal12}
\end{align}
The term $ \sim \; F_{\mu \dot{5}}^0 \delta_ 0  
 \left(  J^{(\varphi ) \, \mu} - \frac{1}{2}
J^{(\chi  ) \, \mu}   \right) $
 in (\ref{dtotal12}) cancels the corresponding  term
in eq. (\ref{dtotal11}).
Also the  term in the last line cancels the first two terms of eq. (\ref{dtotal11}) 
along with $\delta _2$ variations of the gravitino $\psi _{\mu}$ 
and the fermion $\chi $ kinetic terms  
occuring within $\mathcal{L}_0 + \mathcal{L}_b $,
 that is 
$\delta _2 \left( \epsilon^{\mu \nu \rho \sigma} \bar{\psi }_{\mu}
\bar{\sigma}_{\nu} D_{\rho}\psi _{\sigma} -
i  {e^{(5)}} \Delta_{(5)}   \bar{\chi } \bar{\sigma}^{\mu}D_{\mu}\chi \right)$. 
Actually that was the reason nonvanishing 
variations $\delta _2$ had to be introduced for the $\chi$ and $\psi_\mu$ fields.
However not all of the terms
in $\delta _s( \mathcal{L}_0 + \mathcal{L}_b 
+ \Delta _1 \mathcal{L} + \Delta _2 \mathcal{L}) $ are completely cancelled. 
Among those terms that survive is the $ \delta _2 $ variation of $\Delta _1\mathcal{L}$ 
in (\ref{dtotal3}) which  reveals an interesting 
feature that needs be discussed. In fact
\bea
\delta _2 (\Delta _1\mathcal{L}) 
 = -\frac{1}{2} {e^{(5)}} \Delta_{(5)}   
J_{\mu} \delta _2 \psi ^{\mu}
+ \delta _2 \left(-\frac{1}{2} {e^{(5)}} \Delta_{(5)}   
J_{\mu} \right) \psi ^{\mu} + h.c.
\label{dtotal4}
\eea
and the first term in (\ref{dtotal4}), 
after some straightforward algebra, is brought into the form 
\be
- \frac{i}{2\sqrt{2}} 
{e^{(5)}} {\Delta_{(5)}}^2   
\left[
\left( \chi \sigma^{\mu} \bar{\sigma}^{\nu} \varepsilon \right)J_{\mu}^{(\varphi )}
\partial_{\nu} \varphi ^{*} -  \frac{\sqrt{2}}{2} 
\chi \sigma^{\mu} \bar{\sigma}^{\rho}\sigma _{\mu \nu} \varepsilon J^{(\chi ) \nu} 
\partial_{\rho} \varphi ^{*} \right]  + h.c. 
\label{dtotal5} 
\ee
due to the variation $\delta _2\psi _{\mu}$ (see eq. (\ref{coco})). 
Since $i \sqrt{2}\bar{\sigma}^{\nu} \varepsilon \partial_{\nu} \varphi ^{*} $ 
is actually $\delta _0 \bar{\chi} $ we have from 
the expression (\ref{dtotal5}) a contribution $ -\frac{1}{4}
{e^{(5)}} {\Delta_{(5)}}^2   \chi \sigma^{\mu} \left( \delta _0 \bar{\chi } \right)
 \left( J_{\mu}^{(\varphi) } + \frac{1}{4}J_{\mu}^{(\chi)}  \right) + h.c.$
Due to the appearance of this we need to add a new term in the Lagrangian 
which includes, among others, the aforementioned contribution that is
\be
\Delta_3 \mathcal{L} =  \frac{1}{4}  {e^{(5)}} {\Delta_{(5)}}^2    
 J_{\mu}^{(\chi) } \left( J^{(\varphi) \, \mu} + \frac{1}{4}J^{(\chi)  \, \mu} \right)
+ \dotsb
\label{delta3}
\ee
In (\ref{delta3}) the ellipsis denote additional terms.
The terms in (\ref{delta3}) are not new. In ref. \cite{rattazi} such terms do appear 
in the derived Lagrangian completing previous 
derivations, \cite{riotto}. In that work the $\delta ^2(x^5)$ terms complete a perfect 
square (see eq. (3.39) of that paper). This is not the case in our approach. 
However our results at this stage can not be directly compared to those of \cite{rattazi}, 
due to the nontrivial conformal factor existing in eq. (3.39) of the aforementioned paper. 
The relation between the two approaches will be discussed later on.

The above results are easily extended in the case that the original brane action 
has the structure of a general $\sigma$-model (\cite{wittenbagger}), 
\ba
{\mathcal{L}}_{b} = &-& {e^{(5)}} \Delta_{(5)}  \Big[  
K_{ij^*} D_{\mu} \varphi  ^i D^{\mu} \varphi  ^{*j}
+ ( \frac{i}{2} \;K _{ij^*} \chi ^i \sigma ^{\mu} D_{\mu} \bar{\chi} ^j \;+\; h.c )
 \nonumber \\
&+& \frac{1}{2} \; 
( D_i D_j W \; \chi^{i}\chi^{j} \;+\; h.c. )
+ \; K^{ij^*}D_i W \; D_{j^*}W^*  
- \frac{1}{4} R_{i j^* k l^*}\; \chi^i \chi^k \ \bar{\chi}^j \bar{\chi}^l
\Big]
\label{actionflat}
\ea
where $K_{ij^*}$ is the K\"{a}hler metric. In this equation $D_{\mu} \bar{\chi}^j$ is covariant 
under both spacetime and  K\"{a}hler transformations. 
The superpotential and Yukawa terms,   
are also included. In the flat case $D_i W =\partial_i W  $, 
$D_i D_j W = \partial_i \partial_j W - \Gamma_{ij}^{k}\;\partial _k W  $. Later when considering the curved case it turns out that these include additional terms so that they are covariant with respect to the K\"{a}hler function $K$ as well. 
 
The coupling of the 
brane fields to the gauge and the gaugino fields propagating in the bulk is 
known from the flat case, see \cite{peskin}, so that here we will only outline the steps we follow. 
As  can be seen from (\ref{eventrans}) the transformation of  $\lambda ^{a}$, stemming 
from the five dimensions is not exactly that of a gaugino. 
The extra variation requires the addition to the Lagrangian 
of a term
$\;g\; \Delta_{(5)}\;  
D^{(a)} f_x ^ a \partial _{5} \phi ^ x
\;$
while it is known that the variation of the gaugino-fermions Yukawa terms, given by 
$  \Delta_{(5)}\;  \left( \; - ig \sqrt{2} D^{(a)}, _{j^*}  \bar{\chi}^ j \bar{\lambda}^a + h.c. \right)$,  
requires the modification of the gaugino transformation rule by adding a term
$\;
\delta '_{\varepsilon }\lambda ^a = -i g \Delta_{(5)}\;  
D^{(a)}\varepsilon  
\;$
and supersymmetry invariance is finally restored by adding the term
\be
- \; \frac{g^2}{2} {\Delta_{(5)}}^2 \; D^{(a)} D^{(a)}.
\ee
These are the generalizations of the results reached in \cite{peskin} when the 
manifold of the scalar bulk fields is curved.

As far as the presence of the superpotential is concerned, we already know from the flat case 
it modifies the fermions supersymmetry  transformation law according to
\be
\delta^{\prime }_{\varepsilon } \chi^{i} =
 - \sqrt{2}\; K^{ij^*} D_{j^*}W^* \varepsilon \, , 
\ee
The extra variation of the fermion fields applied to the 
coupling of the N\"{o}ther current with the gravitino field $\sim J^{\mu} \psi_{\mu}$, 
see eq. (\ref{first}), leads to modification 
of the gravitino transformation law as
$\;
\delta^{\prime}_{\varepsilon } \psi_{\mu} = i \Delta_{(5)}\;  
 W \sigma_{\mu}\bar{\varepsilon} \, ,
\;$
and the addition to the Lagrangian of the term
\be
{ \mathcal{L}}^{\prime }  =  
e^{(5)} \; \Delta_{(5)}\; \left[
W^* \psi_{\mu} \sigma ^{\mu \nu} \psi_{\nu} + 
W \bar{\psi}_{\mu} \bar{\sigma} ^{\mu \nu} \bar{\psi}_{\nu} \right] \, .
\ee
We can see in turn that its variation, due to $\;   \delta^{\prime}_{\varepsilon } \psi_{\mu} \;$ 
above and the supersymmetry transformation for $\;e^{(4)}\;$, see eq. (\ref{eventrans}),  is
\be
\delta^{\prime }_{\varepsilon } { \mathcal{L}}^{\prime}  
= -3 \Delta_{(5)}\;  
\delta    \left(   {e^{(5)}}  \Delta_{(5)}\;  \right) \mid W \mid ^2
\ee
which is cancelled by the addition of the known $\mid W \mid ^2$ term of the 
supergravity potential which however in our case it appears  
multiplied by ${\Delta_{(5)}}^2$. Variations of the potential terms 
$\;K^{ij^*}D_i W \; D_{j^*}W^*  \;$
are cancelled by the Yukawa terms $\; \sim D_i D_j W \chi^{i}\chi^{j} + h.c. \;$ and those  
of the $\mid W \mid ^2$ require the appearence of terms $\sim  D_i W \chi^i \sigma^{\mu} \bar{\psi}_{\mu} + h.c.$ in the Lagrangian for their cancellation. 
This procedure can be continued and in the following steps the wellknown K\"{a}hlerian exponents 
$e^{\;K/2}$ appearing in the ordinary 4 - D supergravity start showing up accompanying each power of the superpotential $\; W \;$, or derivative of it, both in the Lagrangian and the transformation laws. 
However the K\"{a}hler function $\;K\;$ in the exponent appears multiplied by 
$\; \Delta_{(5)}\; =\; e^5_{\dot{5}} \; \delta (x^5)\;$ as shown in the Lagrangian given below. In conjuction with this we point out 
that the covariant derivatives of the superpotential $\;W\;$ are also found to depend on the 
K\"{a}hler function  through the combination $\;\Delta_{(5)}\;K \;$, rather than  $\;K\;$ itself,  
so that K\"{a}hler invariance is indeed maintained. 

Summarizing, the interactions of a set of  chiral
multiplets localized on the brane designated by the index $i$, with the bulk gravity 
and gauge fields are found to be 
{\allowdisplaybreaks{
\begin{align} 
{\mathcal{L}}^{(4)} &= 
e^{(5)} \Delta_{(5)}\;     \Big[-
 K_{ij^*} D_{\mu} \varphi  ^i D^{\mu} \varphi  ^{*j}
- ( \; \frac{i}{2} K_{ij^*} \chi ^i \sigma ^{\mu} D_{\mu} \bar{\chi} ^j + h.c ) 
 - ig \sqrt{2} (\; D^{(a)}, _{j^*}   \bar{\chi}^ j \bar{\lambda}^a - h.c. \;)
\nonumber  \\
&- \frac{g}{2}D^{(a)} (\; \psi_{\mu} \sigma^{\mu} \bar{ \lambda} ^{a} 
- \bar{\psi}_{\mu} \bar{\sigma} ^{\mu} \lambda ^a  \;) 
- \frac{1}{\sqrt{2}}K_{ij^*} (\; D_{\mu} \varphi  ^{*j} \chi ^i
\sigma ^{\nu} \bar{\sigma} ^{\mu} \psi_{\nu} + D_{\mu} \varphi  ^{i} 
\bar{\chi} ^j
\bar{\sigma} ^{\nu} \sigma ^{\mu} \bar{\psi}_{\nu} 
 \; ) \nonumber \\
 &+ \frac{i}{4}
 E^{\mu \nu \rho \sigma} (\; J_{\sigma}^{(\chi )} - J_{\sigma}^{(\varphi)} \;)
\psi_{\mu} \sigma_{\nu} \bar{\psi}_{\rho} 
+  \frac{1}{4} J_{\sigma}^{(\chi )} 
\psi_{\mu} \sigma^{\sigma} \bar{\psi}^{\mu} +
\frac{1}{4} \Delta_{(5)}\;
 J^{(\chi)\, \mu } \; ( J^{(\varphi)}_{\mu} +  \frac{1}{4}J^{(\chi)}_{\mu} ) 
\nonumber \\
&- \frac{1}{8}\; R_{i{j^*}k{l^*}} \; 
\chi^i \sigma^{\mu} \bar{\chi} ^{j} \; \chi^k \sigma_{\mu} {\bar{\chi}}^{\,l} \; 
- \frac{1}{4}\; ( J^{(\varphi)}_{\mu} -  \frac{1}{2} J^{(\chi)}_{\mu} ) \;
\lambda^a \sigma^\mu {\bar{\lambda}}^a
\nonumber \\
&+ \frac{1}{\sqrt{6}}\; ( - J^{(\varphi ) \, \mu} + \frac{1}{2}
J^{(\chi  ) \, \mu}   )\; F_{\mu \dot{5}}^0 
- \frac{1}{2} g ^2
\Delta_{(5)}\;
\; D^{(a)} D^{(a)} 
+ g D^{(a)} f_x ^ a \partial _{5} \phi ^ x 
\nonumber \\
&- 
 e^{ \Delta_{(5)}\;K/2} ( \;
W^* \psi_{\mu} \sigma ^{\mu \nu} \psi_{\nu}  
+ \frac{i}{\sqrt{2}} D_i W \chi^i \sigma^{\mu} \bar{\psi}_{\mu} 
+ \frac{1}{2} D_i D_j W \chi^{i}\chi^{j} + h.c. \;) \nonumber \\
&-   e^{\Delta_{(5)}\;K} (\;  
K^{ij^*}\; D_i W \; D_{j^*}W^*  - 3 \Delta_{(5)}\; | W |^2 \;)
\Big] +
\dotsb 
\label{sug4}
\end{align}
}}
where in the general case
\be
J_{\mu}^{(\varphi) }= -i \left( K_{i} 
\partial _{\mu} \varphi ^{i } - K_{i^*} \partial_{\mu}
  \varphi^{* i}
\right )
 \,\,, \quad
J_{\mu}^{(\chi) }= K_{ij^*} \chi^{i} \sigma_{\mu} {\bar \chi }^{j} \, . 
\ee
The ellipsis in (\ref{sug4}) stand for couplings of the brane fields with the radion multiplet, which is even, and other even combinations of odd fields which are not presented here. 
 The prefactor $e^{(5)}\;\Delta_{(5)} $ in the Lagrangian above provides $e^{(4)}$ upon integration with respect $x^5$. Note that the terms 
$\;D^{(a)}D^{(a)}$, $J^{(x)}(\cdots)$, $\mid W \mid ^2$ and the exponents involving the 
K\"{a}hler function appear multiplied by an extra power of $\Delta_{(5)}$ whose argument can be put to zero, due to the overall $\Delta_{(5)}$ multiplying the Lagrangian, which is proportional to 
$\; \delta(x^5)\;$. Since $ \int_{-\pi R}^{\pi R} dx^5 e_5^{\dot 5} \Delta_{(5)}(x^5) = 1$ and 
$ \int_{-\pi R}^{\pi R} dx^5 e_5^{\dot 5} = L $ is the "volume" of the 
fifth dimension, we are tempted to   
interpret  $\;\Delta_{(5)}(0) \simeq 1/L \equiv M_L\;$. Replacing then $\;\Delta_{(5)}(0) \;$ 
by $M_L$ and reestablishing units we find that $M_L$ enters in our formulae only through the ratio $\;M_5^3/M_L\;$ where $M_5$ is related to the 
5 - D gravitational coupling through 
$\;k_{(5)}^2\;=\; 1 / M_5^{3}\;$.  The 4 - D gravitational constant is 
$\; k_{(4)}^2= k_{(5)}^2 / L \;$ and the aforementioned ratio is related to the Planck scale via  $ \; {M_5^3} / {M_L} = M_{Planck}^2 \;$. 
In doing all this the gravitino, gauge boson and gaugino, as well as the five dimensional gauge coupling should scale appropriately as $\; \psi_\mu = L^{-1/2} {\hat {\psi}}_\mu \;$, 
$\; {A^{(a)}_\mu} = L^{-1/2} {\hat {A}}^{(a)}_\mu \;$, 
$\; {\lambda^{(a)}_\mu} = L^{-1/2} {\hat {\lambda}}^{(a)}_\mu \;$ and 
$\; g = L^{1/2} g_{(4)}\;$, as dictated by the kinetic terms of these fields, in order for them to have the right normalization and the appropriate dimensions in four dimensions.  
It then turns that with this interpretation the terms in  
(\ref{sug4}) are exactly those encountered in the ordinary 4 - D supergravity involving the interactions of the chiral fields $\varphi^i, \chi^i$ among themselves and their interactions with the gravity and gauge multiplets. Exception to it are additional terms where bulk fields, involving 
$\;F_{\mu \dot{5}}^0, \partial _{5} \phi ^ x $, the radion multiplet etc.,  
interact with the multiplets on the brane. This rather rough qualitative argument is only used to show the correctness of our results. In a decent mathematical way this can be seen after replacing the bulk fields which interact with the brane chiral multiplets by their classical equations of motion as was first done in the model studied in \cite{peskin}. 
This is the case for instance with the D - terms which complete a perfect square, as in the flat case \cite{peskin}, involving the 
derivative $ \partial _{5} \phi ^ x  $. Eliminating this by its classical equation of motion 
results to the ordinary four dimensional D - terms \cite{peskin,riotto}. The importance of the 
$\Delta_{(5)}(0)$ terms at the quantum level has been discussed in \cite{peskin,rattazi,riotto}.

Before continuing two remarks are in order. 
In our approach  we have not 
considered so far gauging of the R-symmetry of the five-dimensional supergravity 
and as a consequence there is no potential on the brane stemming from the coupling 
to the bulk fields \cite{gunaidin,recent}. 
Also the cubic constraint (\ref{solcon}) leads to a $D=4$, 
$N=1$, Yang-Mills supergravity with a gauge field kinetic function $f_{ab}$ 
proportional to $\delta _{ab}$.

Finally in order to make contact with the results reached in \cite{rattazi} we have to express the gravitational part in the unrescaled Weyl basis. This can be accomplished by performing a  Weyl transformation in the four dimensional part of the metric 
$ e_{\mu}^m = \omega \; \tilde {e}_{\mu}^m $ accompanied by the appropriate 
rescalings of the fermionic fields $ \psi _{\mu} = \omega ^{-1/2} \tilde {\psi} _{\mu} \,\,\, , 
\chi   = \omega ^{1/2} \; \tilde {\chi }  $. In order to simplify the discussion we shall limit ourselves to the case of one chiral multiplet on the brane. 
Under this trasformation the gravity action becomes 
\be
-\frac{1}{2}e^{(5)}R^{(5)} =  \tilde {e}^{(5)}
\left( \; \frac{\Omega}{6} \; \tilde {R}^{(5)} -
\frac{1}{4\Omega } \; \Omega,_{\mu} \Omega,^{\mu} + \cdots \right)
\label{weyl}
\ee
which coincides with the corresponding term in \cite{rattazi} if we chose 
$$
\omega ^2 = - \frac{1}{3} \Omega \,, \,\quad \, \Omega = -3 + \Delta_{(5)} \mid \varphi \mid ^2.
$$

Note that in the above relation the non-trivial part of the $\Omega$ acts only on the brane.
The second term in (\ref{weyl}) combined with the scalar field kinetic term  gives 
\be
- \Omega,_{\varphi \varphi ^*} \partial _{\mu}  \varphi  \partial ^{\mu} \varphi ^* + 
\frac{1}{4 \Omega} \; {\hat J}^{(\varphi)}_{\mu} {\hat J}^{(\varphi) \mu} 
\label{resc}
\ee
if the K\"{a}hler function $\;K\;$ is related to $\; \Omega \;$ through 
$\; \Delta_{(5)}\;K = -3 \; ln ( - \frac{\Omega }{3} )\;$. 
The hated currents  $\; {\hat J}^{(\varphi)}_{\mu}, {\hat J}^{(\chi)}_{\mu} \;$ are defined as 
$\; {\hat J}^{(\varphi)}_{\mu} = -i ( \Omega_{\varphi} \partial_{\mu} \varphi  - h.c. ) \;$ and 
$\; {\hat J}^{(\chi)}_{\mu} =   \Omega_{\varphi \varphi^*} 
{\tilde {\chi}} \sigma_\mu  {\bar {\tilde {\chi}}} \;$ respectively. Now in order to bring the fermion kinetic term to a form proportional to  
$\; \Omega,_{\varphi \varphi ^*}\;$, as in the scalars kinetic terms above,   
we have to shift the gravitino on the brane as  
$\tilde { \psi } _{\mu} = \hat {\psi} + 
i \;  \frac{\; \; \Omega_{\varphi^*}}{{\sqrt 2} \Omega}  \;  \sigma_{\mu} \, {\bar {\tilde { \chi}}} $. 
This except of putting the fermion kinetic term to its canonical form, in the above sense, produces a term 
$\;  \tilde{e}^{(5)} \frac{1}{2 \Omega}{\hat J}^{(\varphi)}_{\mu} {\hat J}^{(\chi ) \,\, \mu}\; $. Such terms are also generated after the gravitino shift from the rescaled coupling of 
the N{\" o}ether current with the gravitino yielding 
$\; - \tilde{e}^{(5)} \frac{1}{\Omega} {\hat J}^{(\varphi)}_{\mu} {\hat J}^{(\chi ) \,\, \mu}\; $. 
A third source of similar terms are those given by 
$$
-e^{(5)} \frac{i}{4} \Delta_{(5)} \left[\Delta_{(5)} K,_{\varphi \varphi ^*}
( K,_{\varphi} \partial _{\mu}\varphi  - K,_{\varphi^*} \partial _{\mu} \varphi^*  ) - 
2(K,_{ \varphi \varphi \varphi ^*} \partial _{\mu }\varphi 
- K,_{\varphi^* \varphi^* \varphi } \partial _{\mu}{\varphi^* } ) \right]{ \chi} \sigma^{\mu} \bar{ {\chi} }
$$
where the first stems from $\; \sim J^{(\varphi) \mu} J^{(\chi)}_\mu \;$
and the second  from the K{\"a}hler covariantization of the fermion kinetic terms 
in (\ref{sug4}). 
These terms after the rescaling yield 
$ \tilde{e}^{(5)} \frac{1}{4\Omega} {\hat J}^{(\varphi)}_{\mu} {\hat J}^{(\chi) \mu }\;$. 
Adding these we find that their total contribution to the rescaled Lagrangian is 
$ \; - \tilde{e}^{(5)} \frac{1}{4\Omega} {\hat J}^{(\varphi)}_{\mu} {\hat J}^{(\chi) \mu }\;$. 

Finally from the 
$ \; { J}^{(\chi)}_{\mu} { J}^{(\chi ) \mu}\;$ and the K\"{a}hler curvature term in 
(\ref{sug4}) we get, after  performing the Weyl rescalings,  
$ \tilde{e}^{(5)} \frac{1}{16 \Omega} {\hat J}^{(\chi)}_{\mu} {\hat J}^{(\chi ) \mu}\;$. Collecting then the 
$\;{\hat J}^{(\varphi)} \cdot {\hat J}^{(\chi )}, \; {\hat J}^{(\chi)} \cdot {\hat J}^{(\chi )}\;$ 
terms 
and the  $\;{\hat J}^{(\varphi)} \cdot {\hat J}^{(\varphi )}\;$ term in (\ref{resc}) we arrive at the result 
$$
 \tilde{e}^{(5)} \frac{1}{4\Omega} \left[ {\hat J}^{(\varphi)}- \frac{1}{2}{\hat J}^{(\chi )} \right]^2
$$
completing a perfect square of the matter currents.
This term combined with those that are linear and bilinear  in $ F_{\mu \dot{5}} $ yields a  perfect square given by
\be
\tilde{e}^{(5)} \frac{3}{2\Omega} \left[ \hat {F}_{\mu \dot{5}} - \frac{1}{\sqrt{6}}\left(
 {\hat J}^{(\varphi)}_\mu- \frac{1}{2}{\hat J}^{(\chi )}_\mu \right) \right]^2 
\label{ff}
\ee
where $\hat {F}_{\mu \dot{5}} \equiv (- \Omega / 3 ) F_{\mu {5}} \;e^{5}_{\dot 5}$. 

We therefore see that by a Weyl rescaling and collecting the necessary terms we are able to derive the Lagrangian terms given by eq. (3.39) in \cite{rattazi}. 
However in eq. (\ref{ff}) $\hat {F}_{\mu \dot{5}}$ appears instead of
${F}_{\mu \dot{5}}$ found in that reference.
This apparently small difference has a rather major impact on the 4-D effective supergravity  Lagrangian since one does not get a regular current-current interaction after integrating
out the graviphoton field. 
The source of the discrepancy can be sought in
the on-shell method employed in this work, in conjuction with the order, in the 5-D
gravitational constant $k_{(5)}$,  the couplings of the
graviphoton to the brane fields have been derived. In our own frame and in the 
on-shell procedure except the existing ${F}_{\mu \dot{5}}^2$ kinetic term, which is of zeroth order, the interaction  ${F}_{\mu \dot{5}} \; J^{\mu}$ was derived up to 
order ${\cal{O}} (k_{(5)})$. 
If we continue carrying out Noether's procedure
higher order terms will be collected involving ${F}_{\mu \dot{5}}^2$ and
${F}_{\mu \dot{5}} \; J^{\mu}$. In order to reconcile (\ref{ff}) with the findings of ref. \cite{rattazi} these terms should exponentiate as  
$ e^{ \;  \frac{2}{3} \; \Delta_{(5)} K  } \; {F}_{\mu \dot{5}}^2$ and
$e^{\;  \frac{1}{3} \; \Delta_{(5)} K  } \; {F}_{\mu \dot{5}} \;
J^{\mu}$  respectively. Note that the exponent $ \Delta_{(5)} K  $ as well as the singular part of $\Omega$ are both of order ${\cal{O}} ( k_{(5)}^2 )$. This may not be unfeasable. In fact  exponentials of this sort, involving the
K\"{a}hler function $\; K \;$, do indeed appear in the action of the
supersymmetric sigma model when we localize the global supersymmetry.
The appearance of these exponentials is equivalent to saying that the field 
$\;{F}_{\mu \dot{5}}$ is renormalized to
$e^{   \frac{1}{3} \; \Delta_{(5)} K  } \;{F}_{\mu \dot{5}} $, or the same 
$( \;-3/\Omega \;) \;{F}_{\mu \dot{5}}, $ having as effect the replacement of 
$\hat {F}_{\mu \dot{5}}$ in eq. (\ref{ff})  by  $ {F}_{\mu \dot{5}}$, obtaining thus complete agreement with \cite{rattazi} to all orders in $k_{(5)}^2$.  
Having in mind that the the coupling of  ${F}_{\mu \dot{5}}$ to matter oughts to be of the form presented in \cite{rattazi}, for it leads to the correct current-current interaction in the effective 4-D Lagrangian, we argue that Noether's procedure 
oughts to yield the above renormalization for ${F}_{\mu \dot{5}}$ in the sense outlined previously.  
In order to check if this is indeed the case higher order interactions, in the gravitational constant $k_{(5)}$, of the brane fields with the five-dimensional gravity multiplet have to be derived in the on-shell scheme we have adopted. This rather complicated task, along with the derivation of additional terms coupling the brane fields to the radion multiplet, and other even combination of odd fields, which complete the Lagrangian given by (\ref{sug4}), will 
appear in a future publication.

\section{Discussion}

In the context of $D=5$, $N=2$, Yang-Mills Supergavity compactified on 
$S^1/Z_2$  we  consider the supersymmetric coupling of matter fields propagating 
on the brane at $x^5 = 0$. Working in the on-shell scheme we have derived the terms of 
the brane action which are relevant for studying the mechanisms 
of supersymmetry  and  gauge symmetry breaking.  The omitted radion multiplet couplings, as well as other couplings to the brane fields, can be  
derived, if desired, using the N\"{o}ther procedure which we followed in this paper. 
The complete  brane action including these terms and the mechanisms of supersymmetry and the gauge symmetry breaking  
in particular unified models, in which both Gravity and Gauge forces propagate in the bulk, 
will be the issue of a forthcoming publication.
\\
\\
\noindent
{\bf Acknowledgements} \\ 
The authors acknowledge partial financial support
from the Athens University special account for research and PYTHAGORAS grant.   
\noindent A.B.L. acknowledges also support from HPRN-CT-2000-00148
and HPRN-CT-2000-00149 programmes. 

\end{document}